\documentclass{ws-ijmpe}

\begin{document}

\markboth{Terence Tarnowsky}{Energy Dependence of Short and Long-Range Multiplicity Correlations in Au+Au Collisions from STAR}

\catchline{}{}{}{}{}

\title{ENERGY DEPENDENCE OF SHORT AND LONG-RANGE MULTIPLICITY CORRELATIONS IN AU+AU COLLISIONS FROM STAR}

\author{\footnotesize TERENCE TARNOWSKY (FOR THE STAR COLLABORATION)}

\address{Physics Department, Purdue University, 525 Northwestern Ave\\
West Lafayette, IN 47907,
USA\\
tjt@physics.purdue.edu}

%
%
\maketitle

\begin{history}
\received{(received date)}
\revised{(revised date)}
\end{history}

\begin{abstract}
A general overview of the measurement of long-range multiplicity correlations measured by the STAR experiment in Au+Au collisions at RHIC is presented. The presence of long-range correlations can provide insight into the early stages, and the type of matter produced in, these collisions. These measurements have been made in Au+Au collisions at $\sqrt{s_{NN}}$ = 200 and 62.4 GeV. These results indicate a relatively large long-range correlation is produced in Au+Au collisions compared to a {\it pp} baseline at $\sqrt{s_{NN}}$ = 200 GeV. A weaker long-range correlation is seen as a function of incident energy. Further, comparison of the onset of the long-range correlation to the calculated percolation density parameter at $\sqrt{s_{NN}}$ = 200 GeV is presented. 
\end{abstract}

\section{Introduction}

The study of particle correlations has been suggested as a means to search for a possible phase transition in ultra-relativistic heavy ion collisions.\cite{1,2,3} These particle correlations have the potential to probe the early stages of heavy ion collisions. If a deconfined phase of quarks and gluons exists in these collisions, the presence of these partonic degrees of freedom could directly influence the results of a correlation measurement. One such measurement is the study of long-range multiplicity correlations that extend over a range in pseudorapidity ($\eta$). A linear relationship has been experimentally shown in hadron-hadron collisions relating the multiplicity in the forward direction ($N_{f}$) to the average multiplicity in the backward direction ($N_{b}$).\cite{4} 

\begin{equation}\label{linear} 
<N_{b}(N_{f})> = a + bN_{f} 
\end{equation}

The coefficient $b$ is referred to as the correlation coefficient. The value, $a$, is a measure of uncorrelated particles. In the definition provided in ~``Eq.~(\ref{linear})'' the maximum value of $b$ is one, corresponding to complete correlation between the produced particles. For $b = 0$, all particles are uncorrelated. This relationship can be expressed in terms of the following expectation value:\cite{5}

\begin{equation}\label{b}
b = \frac{<N_{f}N_{b}>-<N_{f}><N_{b}>}{<N_{f}^{2}>-<N_{f}>^{2}} = \frac{D_{bf}^{2}}{D_{ff}^{2}}
\end{equation}
where $D_{bf}^{2}$ and $D_{ff}^{2}$ are the backward-forward and forward-forward dispersions, respectively. This result is exact and model independent.\cite{6} All quantities in ~``Eq.~(\ref{b})'' are directly measurable. The large pseudorapidity and angular coverage of the STAR experiment at RHIC provides an ideal means to measure particle correlations with precision.\cite{7}

In phenomenological particle production models, the hadronization of color strings formed by partons in the collision account for the produced soft particles. In the Dual Parton Model (DPM), at low energies these strings are formed between valence quarks.\cite{5} At higher energies, there are additional strings formed due to the increasing contribution from sea quarks and gluons. In the DPM, this production mechanism is predicted to produce long-range multiplicity correlations in pseudorapidity. A long-range correlation could be an indication of multiple elementary inelastic collisions in the initial stages of an A+A collision.\cite{5} Additionally, the Color Glass Condensate (CGC) model predicts long-range correlations.\cite{8} The CGC model is a description of high density gluonic matter at small-x. In the CGC model, the incoming, Lorentz contracted nuclei are envisioned as two sheets of colliding colored glass.\cite{9} In the initial stage of the collision, the color fields are oriented transverse to the beam direction. After the collision these color fields realign longitudinally. This produced state is referred to as the Glasma, an intermediate state between the CGC and Quark Gluon Plasma (QGP).\cite{10} These longitudinal fields provide the origin of the long-range correlation and are similar to the strings in the DPM.

%

\section{Data Analysis}

\begin{figure}
\centering
\includegraphics[width=4in]{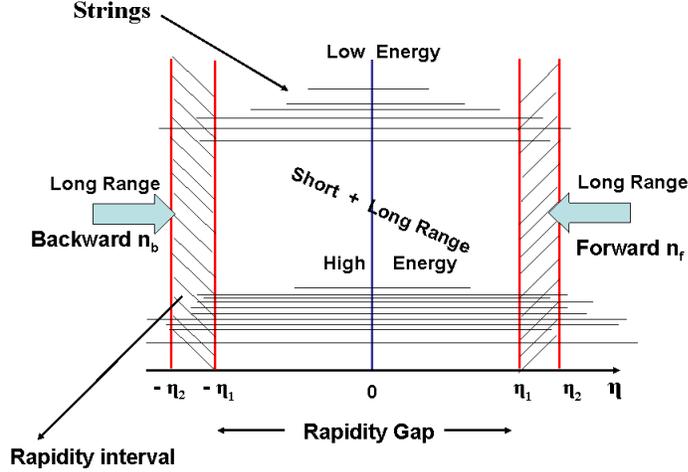}
\caption{\footnotesize{Schematic diagram of the measurement of a forward-backward correlation.}}
\label{fig1}
\end{figure}

The data utilized for this analysis are from year 2001 (Run II) $\sqrt{s_{NN}}$ = 200 GeV Au+Au and year 2004 $\sqrt{s_{NN}}$ = 62.4 GeV collisions at the Relativistic Heavy Ion Collider (RHIC), as measured by the STAR (Solenoidal Tracker at RHIC) experiment.\cite{7} The main tracking detector at STAR is the Time Projection Chamber (TPC).\cite{11} The TPC is located inside a solenoidal magnet generating a constant, longitudinal magnetic field. For this analysis, data was acquired at the maximum field strength of 0.5 T. All charged particles in the TPC pseudorapidity range $0.0 < |\eta| < 1.0$ and with $p_{T} > 0.15$ GeV/c were considered. This $\eta$ range was subdivided into forward and backward measurement intervals of total width 0.2$\eta$. The forward-backward correlations were measured symmetrically about $\eta = 0$ with varying $\eta$ gaps (measured from the center of each bin), $\Delta\eta =$ 0.2, 0.4, 0.6, 0.8, 1.0, 1.2, 1.4, 1.6, and 1.8 units in pseudorapidity. Fig \ref{fig1} depicts a schematic representation of the forward-backward correlation and its measurement. The collision events were part of the minimum bias dataset. The minimum bias collision centrality was determined by an offline cut on the TPC charged particle multiplicity within the range $|\eta| < 0.5$. The centralities used in this analysis account for 0-10, 10-20, 20-30, 30-40, 40-50, 50-60, 60-70, and 70-80\% of the total hadronic cross section. An additional offline cut on the longitudinal position of the collision vertex ($v_{z}$) restricted it to within $\pm 30$ cm from $z=0$ (center of the TPC). Approximately two million minimum bias events at $\sqrt{s_{NN}}$ = 200 GeV and 0.7 million events at $\sqrt{s_{NN}}$ = 62.4 GeV satisfied these requirements and were used for this analysis. Corrections for detector geometric acceptance and tracking efficiency were carried out using a Monte Carlo event generator and propagating the simulated particles through a GEANT representation of the STAR detector geometry. 

In order to eliminate the effect of statistical impact parameter (centrality) fluctuations on this measurement, each relevant quantity ($\left<N_{f}\right>$, $\left<N_{b}\right>$, $\left<N_{f}\right>^{2}$, and $\left<N_{f}N_{b}\right>$) was obtained on an event-by-event basis as a function of STAR reference multiplicity, $N_{ch}$. A linear fit to $\left<N_{f}\right>$ and $\left<N_{b}\right>$, or a second order polynomial fit to $\left<N_{f}\right>^{2}$ and $\left<N_{f}N_{b}\right>$ was used to extract these quantities as functions of $N_{ch}$. Tracking efficiency and acceptance corrections were applied to each event. These were then used to calculate the backward-forward and forward-forward dispersions, $D_{bf}^{2}$ and $D_{ff}^{2}$, binned according to the STAR centrality definitions and normalized by the total number of events in each bin. An autocorrelation exists if the centrality definition overlaps with the measurement interval in $\eta$ space. Therefore, there are three $\eta$ regions used to determine reference multiplicity: $|\eta| < 0.5$ (for measurements of $\Delta\eta < 1.0$), $0.5 < |\eta| < 1.0$ (for measurements of $\Delta\eta > 1.0$), or $|\eta| < 0.3 + 0.6 < |\eta| < 0.8$ (for measurement of $\Delta\eta = 1.0$). This is possible due to the flat $\eta$ yield of the STAR TPC within the measurement ranges being considered. 

\begin{figure}
\centering
\includegraphics[width=4in]{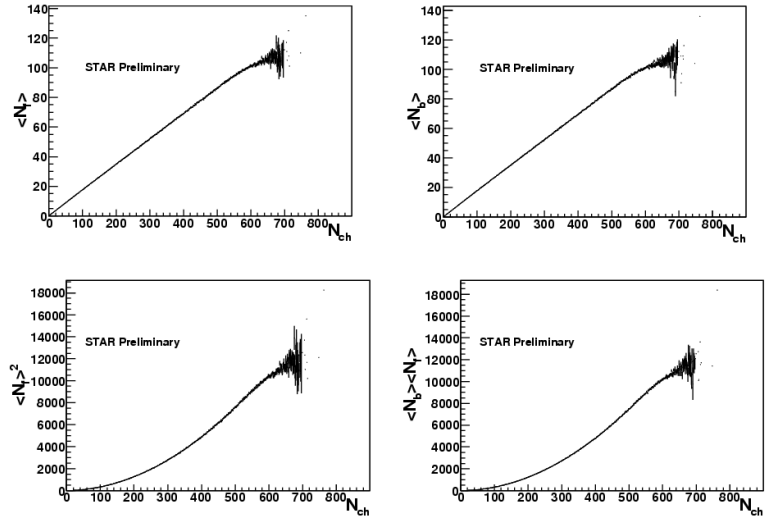}
\caption{\footnotesize{$<N_{f}>$, $<N_{b}>$, $<N_{f}>^{2}$, and $<N_{f}*N_{b}>$ as functions of STAR charged particle reference multiplicity ($N_{ch}$). The quantities are found event-by-event for all values of $N_{ch}$.}}

\label{fig2}
\end{figure}

\section{Results}

Fig. \ref{fig2} presents the mean multiplicity $\left<N_{f}\right>$ and $\left<N_{b}\right>$, in a forward and backward pseudorapidity region along with $\left<N_{f}\right>^{2}$ and $\left<N_{f}N_{b}\right>$ as functions of reference multiplicity, $N_{ch}$. The results are calculated event-by-event for every unit $N_{ch}$. The $\left<N_{f}\right>$ and $\left<N_{b}\right>$ demonstrate a linear dependence over much of the centrality range. Only statistical uncertainties are present and are less than 1\%, except for the highest multiplicities. The results in Fig. \ref{fig2} are uncorrected for tracking efficiency and acceptance. Due to statistical limitations, it is not possible to apply corrections for every value of $N_{ch}$. One value for the correction, calculated for each centrality, is applied to the resultant values $\left<N_{f}\right>, \left<N_{b}\right>, \left<N_{f}\right>^{2}$, and $\left<N_{f}N_{b}\right>$ for every event in that centrality. Therefore, all events falling within a particular centrality have the same correction. 

\begin{figure}
\centering
\includegraphics[width=4in]{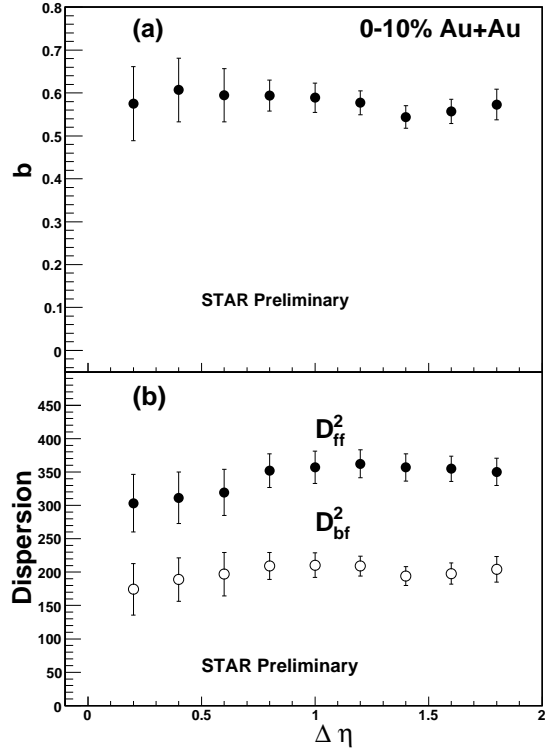}
\caption{\footnotesize{(a) Correlation coefficient, $b$ and (b) Dispersions, $D_{bf}^{2}$ and $D_{ff}^{2}$ as a function of the pseudorapidity gap ($\Delta\eta$) in 0-10\% most central $\sqrt{s_{NN}}$ = 200 GeV Au+Au collisions. If only short-range correlations were present, the expectation is a rapidly falling $b$ value with increasing $\Delta\eta$. The approximately flat behavior of $b$ indicates the presence of a long-range correlation.}}
\label{fig3}
\end{figure}

Fig. \ref{fig3} shows the results for the correlation coefficient $b$ and the backward-forward and forward-forward dispersions, $D_{bf}^{2}$ and $D_{ff}^{2}$, for most central (0-10\%) Au+Au collisions at $\sqrt{s_{NN}}$ = 200 GeV. The correlation is approximately flat as a function of the pseudorapidity gap. It is expected that if only short-range correlations are present (mostly from sources such as cluster formation, jets, or resonance decay), then $b$ would decrease rapidly as a function of $\Delta\eta$.\footnote{Note that there are methods to estimate the short-range contribution and remove it, thereby providing an absolute measure of the ``pure'' long-range component. This is beyond the scope of the current presentation, but is discussed in the following.\cite{13}} It has been argued that the presence of these long-range correlations at larger than $\Delta\eta = 1.0$ cannot be due to final state effects, and are therefore due to the conditions in the initial state.\cite{8} This may be phenomenologically similar to fluctuations larger than the event horizon being created during an inflationary phase in the early universe that later become smaller than the event horizon.\cite{12}

\begin{figure}
\centering
\includegraphics[width=4in]{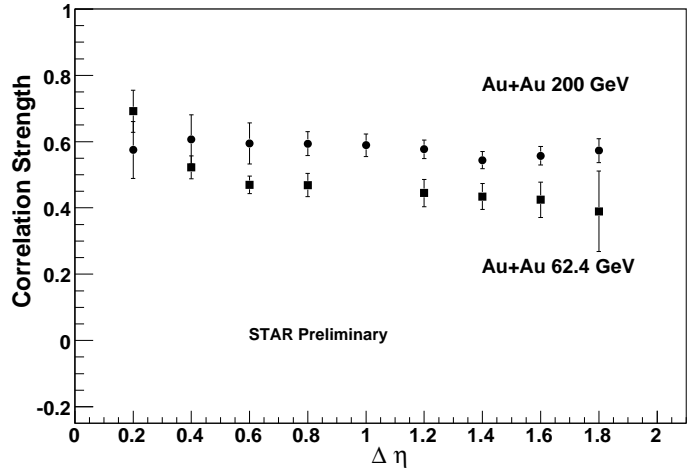}
\caption{\footnotesize{(a) Comparison of the correlation coefficient, $b$, as a function of the pseudorapidity gap ($\Delta\eta$) in 0-10\% most central $\sqrt{s_{NN}}$ = 200 and 62.4 GeV Au+Au collisions. The lower energy data exhibits a similar plateau as the 200 GeV result, but at a lower absolute number. This could indicate a smaller long-range correlation in $\sqrt{s_{NN}}$ = 62.4 GeV Au+Au collisions.}}
\label{fig4}
\end{figure}

A comparison of the correlation coefficient, $b$, in 0-10\% most central $\sqrt{s_{NN}}$ = 200 and 62.4 GeV is shown in Fig. \ref{fig4}. There is a slight difference in the evolution of the correlation at small values of $\Delta\eta$, but for $\Delta\eta > 0.6$ the value of the correlation coefficient is approximately flat as a function of $\Delta\eta$ for the two energies. However, for the $\sqrt{s_{NN}}$ = 62.4 GeV data, the value of $b$ is lower by up to $\approx$ 15\%. It is expected that $b$ has a dependence on incident energy (and atomic number) of the colliding nuclei.\cite{5} Data for Cu+Cu collisions at the same energies are currently under study. 

\begin{figure}
\centering
\includegraphics[width=2.5in]{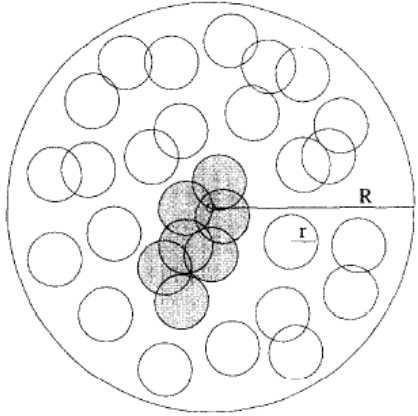}
\caption{\footnotesize{Schematic of cluster formation in 2-D percolation.\protect\cite{15}}}
\label{fig5}
\end{figure}

\begin{figure}
\centering
\includegraphics[width=4in]{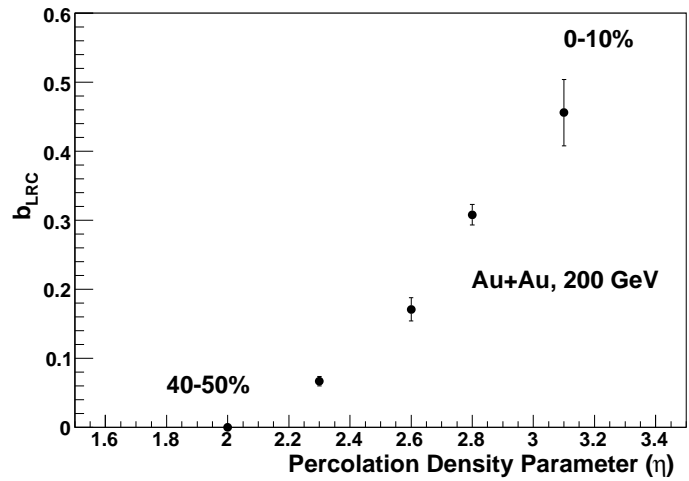}
\caption{\footnotesize{Extracted long-range correlation, $b_{LRC}$, as a function of the percolation density parameter, $\eta$, for Au+Au at $\sqrt{s_{NN}}$ = 200 GeV. At the point where the long-range component fades, $\eta$ is $\approxeq$ 2. Most theoretical predictions for $eta_{c}$ lie below this.}}
\label{fig6}
\end{figure}

The measurement of a long-range correlation demonstrates a coherent phenomenon in the initial state of these heavy ion collisions. It may then be possible to correlate this effect with that of another collective phenomenon, percolation. The particle production mechanism in high-energy heavy ion collisions can be described by the hadronization of color strings stretched between the colliding nuclei.\cite{14} At high energies it is possible for partons from different nucleons to begin to overlap in the plane transverse to the propagation direction of the nuclei, forming clusters (Fig. \ref{fig5}).\cite{15,16} At the percolation threshold, a cluster that spans the entire system is formed. This point can be characterized by the critical percolation density parameter, $\eta_{c}$. $\eta_{c}$ has been predicted theoretically\cite{17,18,19} and the percolation density parameter itself has been measured experimentally in heavy ion collisions by the STAR experiment.\cite{20,21}. It has been shown by these measurements that the critical value, $\eta_{c}$, for the ``local'' prediction ($\eta_{c}$ = 1.72) is exceeded in most centralities of Au+Au collisions at $\sqrt{s_{NN}}$ = 200, while for $\sqrt{s_{NN}}$ = 62.4 Au+Au collisions all centralities lie at or below the critical percolation threshold. A slightly lower value for $\eta_{c}$ is predicted in the infinite (continuum) system. In that case, central Au+Au at $\sqrt{s_{NN}}$ = 62.4 GeV will lie above $\eta_{c}$. It is thus important to try and experimentally constrain $\eta_{c}$. If the presence of long-range correlations and percolation are related, it may be possible to constrain $\eta_{c}$ by examining the long-range correlation as a function of the measured percolation density parameter, shown in Fig. \ref{fig6}. The $b$ shown here is the long-range component only.\cite{13} It is seen that the extracted long-range component of the correlation coefficient $b$ crosses zero in the area around a percolation density parameter value of two. This is higher than most theoretically predicted values for $\eta_{c}$. Though this result is intriguing, further study is required.

\section{Summary}

Results have been presented on the energy dependence of the forward-backward correlation coefficient, $b$, in Au+Au collisions. It is seen that the value of $b$ is approximately flat over a wide range in pseudorapidity gap, $\Delta\eta$, at $\sqrt{s_{NN}}$ = 200 GeV. This indicates a long-range correlation as a function of $\Delta\eta$. Were only short-range correlations present, a rapid decay in the value of $b$ would be expected as $\Delta\eta$ increases. These long-range correlations can be ascribed to multiple elementary inelastic collisions. These are predicted in models such as the Dual Parton Model and the Color Glass Condensate/Glasma phenomenology. The long-range correlation appears to persist at $\sqrt{s_{NN}}$ = 62.4 GeV, though at a lower value than $\sqrt{s_{NN}}$ = 200 GeV. Further investigations are proceeding to determine an experimental range for the critical percolation density parameter, $\eta_{c}$. At the point where the extracted long-range correlation value goes to zero, the value of the percolation density parameter is $\approx$ 2. This is larger than several theoretical predictions for $\eta_{c}$. Additional study is required to characterize this relationship.

\end{document}